\documentclass[a4paper,12pt]{article}
\usepackage{jheppub} 

\newcommand{\mpt}{p\hspace{-0.45em}/ }

\title{3.5-keV X-ray line from nearly-degenerate WIMP dark matter decays
}

\author[a,b,c]{Cheng-Wei~Chiang}
\author[a]{and Toshifumi~Yamada}

\affiliation[a]{Department of Physics and Center for Mathematics and Theoretical Physics, \\
National Central University, Chungli, Taiwan 32001, Republic of China}
\affiliation[b]{Institute of Physics, Academia Sinica, Taipei, Taiwan 11529, Republic of China}
\affiliation[c]{Physics Division, National Center for Theoretical Sciences, \\
Hsinchu, Taiwan 30013, Republic of China}

\abstract{
The unidentified emission line at the energy of $\sim$3.5~keV observed in X-rays from galaxy clusters may originate from a process involving a dark matter particle.
On the other hand, a weakly interacting massive particle (WIMP) has been an attractive dark matter candidate, due to its well-understood thermal production mechanism and its connection to physics at the TeV scale.
In this paper, we pursue the possibility that the 3.5-keV X-ray arises from a late time decay of a WIMP dark matter into another WIMP dark matter, both of which have the mass of $O(100)$~GeV and whose mass splitting is about 3.5~keV.
We focus on the simplest case where there are two Majorana dark matter particles and two charged scalars that couple with a standard model matter particle.
By assuming a hierarchical structure in the couplings of the two dark matter particles and two charged scalars, it is possible to explain the 3.5-keV line and realize the WIMP dark matter scenario at the same time.
Since the effective coupling of the two different Majorana dark matter particles and one photon violates CP symmetry, the model always contains a new source of CP violation, so the model's connection to the physics of electric dipole moments is discussed.
The model's peculiar signatures at the LHC are also studied.  We show the prospect of detecting the charged scalars through a detailed collider simulation.
}


\begin{document} 
\maketitle
\flushbottom

\section{Introduction \label{sec:intro}}

\ \ \ Although existence of dark matter (DM) is firmly believed through observations of its gravitational effects,
 its detailed properties and interactions with the visible sector other than gravity remain largely unknown.
Recently, a faint hint on the nature of DM has been reported by two groups \cite{report1,report2} through observations of the X-ray spectrum from galaxy clusters.
They claim that they have discovered a weak X-ray emission line at the energy of about 3.5~keV that
 cannot be explained by any known atomic electron transition in thermal plasma,
 and that this line is possibly due to the decay of a DM particle, with
 one example being the decay of a 7~keV sterile neutrino into a photon and an active neutrino.
If the emission line originates from a DM particle, we then have an important clue on possible interactions of the DM; namely, this particle couples with the photon in such a way that its interaction involves a 3.5-keV monochromatic photon.
Many studies have been done on the connection between the DM physics and the 3.5-keV X-ray line \cite{studies}.

On the other hand, there has been another hint on the nature of DM from theoretical studies; that is, the so-called weakly interacting massive particle (WIMP) miracle.
If the DM particle that is responsible for the observed DM abundance couples with a standard model (SM) particle through a coupling constant of order 0.1, and if it is a cold relic of thermal plasma in the early Universe (these are the case when the DM particle is a WIMP), then its interaction with the SM sector should involve a particle of $O(100)$~GeV mass, just above the electroweak symmetry breaking scale where new physics is expected to enter.
In addition, if the DM particle has the same new physics origin as the new particle that mediates the interaction between the DM sector and the SM sector, then it is natural to consider that the DM particle itself has the mass of $O(100)$~GeV.

In this paper, we propose a scenario that accommodates both the hint of DM that may be responsible for the 3.5-keV X-ray emission line and the scenario of WIMPs with mass of $O(100)$~GeV.
In the scenario, there are two species of WIMP DM particles whose masses are both $O(100)$~GeV and whose mass difference is about 3.5~keV.
Both DM species are stable at the cosmological time scale.  
The heavier DM particle decays into the lighter one and a 3.5-keV photon with a rate below $\sim 1/t_{cos}$, where $t_{cos}$ denotes the age of the Universe,
 giving rise to the observed monochromatic 3.5-keV X-ray emission line.

Let us now determine the structure of one of the simplest models.
First, we discuss in what cases the decay of a DM particle into another DM particle and a photon is realized.
Scalar DM particles are impossible because the decay of a scalar particle into another scalar particle and a photon is forbidden by angular momentum conservation.
Hence the next possibility is the case with spin-1/2 Majorana fermions.
Majorana particles interact with an \textit{on-shell} photon \textit{only} through the following dipole transition operators \cite{transition dipole}, where $\chi_1$ and $\chi_2$ are different mass eigenstates:
\begin{align}
\frac{1}{\Lambda} \bar{\chi}_2 \sigma_{\mu \nu} \chi_1 \, F^{\mu \nu}, \ \ \ 
\frac{1}{\Lambda} \bar{\chi}_2 \sigma_{\mu \nu} \gamma_5 \chi_1 \, F^{\mu \nu}.
\label{maj}
\end{align}
In fact, it is known that spin-1/2 particles can interact with a photon also through an anapole coupling \cite{anapole}.  But this coupling vanishes identically when the photon is on shell.

Next, we pin down the structure of the interaction between our Majorana DM particles and SM particles that allows the DM particles to be the WIMPs.
If the DM particles interact with the SM sector only through the operators in Eq.~(\ref{maj}) with ${\cal O}(1)$ coefficients, then WIMP miracle would suggest that $\Lambda = O(100)-O(1000)$~GeV.
However, this would give too large a decay rate for the heavier DM particle, far above the inverse of the age of the Universe, leading to the situation where the abundance of the heavier DM at present is extremely suppressed and the photon flux coming from its decay is insufficient to explain the observed 3.5-keV X-ray line.
One simple solution to this difficulty is to consider the following renormalizable interaction Hamiltonian which induces the interactions in Eq.~(\ref{maj}) at one-loop level,
\begin{align}
{\cal H}_{{\rm int}} \ &= \ \lambda^L_{ij} \, \bar{\chi}_i P_L \psi S_j \ + \ \lambda^R_{ij} \, \bar{\chi}_i P_R \psi S_j \ + \ {\rm h.c.} \ \ \ (i,j=1,2)~,
\label{renormalizable}
\end{align}
 where $S_1, S_2$ denote two scalar fields and $\psi$ denotes a spin-1/2 field, both of which have electric charges, and the coupling constants $\lambda_{11}, \lambda_{22}$ are assumed to be of $O(0.1)$ whereas $\lambda_{12}, \lambda_{21}$ are much more suppressed.  
In the following, we assume that the decays of $\chi_i \to \psi S_j$ are kinematically forbidden for the stability of the DM particles.

The right thermal relic abundance of DM particles can be achieved in the following two scenarios.
In the first scenario, $\psi$ is lighter than $\chi_i$s and $S_i$s have the mass of $O(100)$~GeV.  
In this case, $\psi$ is in thermal equilibrium with SM particles through the electroweak interactions when $\chi_i$s go out of thermal equilibrium.  
The fact that $\lambda_{11}, \lambda_{22}$ are $O(0.1)$ and $S_i$s have the mass of $O(100)$~GeV guarantees that the thermal relic abundance of $\chi_i$s fits the observed DM abundance $\Omega_{DM}h^2 \sim 0.1$.
The other possibility is that $S_i$s are lighter than $\chi_i$s and $\psi$ has the mass of $O(100)$~GeV, in which case the thermal relic abundance of $\chi_i$s again can fit the observed DM abundance.
To avoid the presence of a stable charged particle, $\psi$ (if $\psi$ is lighter than the DMs) or $S_i$s (if $S_i$s are lighter than the DMs) should be or eventually decay into SM particles.
We shall adopt the simplest case in which $\psi$ is lighter than the DM particles and is identified as a SM matter field with an electric charge.

On the other hand, the decay rate of the heavier DM species into the lighter one and the photon is controlled by the coupling constants $\lambda_{12}$ and $\lambda_{21}$, and hence the photon flux of the 3.5-keV X-ray line can be explained by taking appropriate values for these constants.

To summarize, the simplest model that can explain the 3.5-keV X-ray line by a WIMP DM decay into another WIMP DM and a photon has two species of spin-1/2 Majorana DM particles that couple with two charged scalar particles of $O(100)$~GeV mass and an electrically-charged SM particle.
The coupling constants for the DM particles have a hierarchical structure that each DM particle dominantly couples with a distinct charged scalar with the strength of $O(0.1)$, and couples with the other charged scalar with a much more suppressed strength.
The $O(0.1)$ coupling constants are responsible for obtaining the right thermal relic abundance of DM particles, as they control the decoupling of DM particles from the thermal bath in the early Universe.
On the other hand, the suppressed coupling constants are responsible for reproducing the observed photon flux of the 3.5-keV X-ray line, as they determine the decay rate of the heavier DM species into the lighter one and the photon.
Phenomenologically important is the fact that the combination of the WIMP DM scenario and the 3.5-keV X-ray line predicts, in the simplest case, the existence of \textit{two} charged scalars that couple with the DM particles and a SM matter field, which is testable in collider experiments.

In this paper, we construct a concrete WIMP DM model based on the above arguments, and derive the coupling constants $\lambda_{ij}$s that reproduce the observed flux of the 3.5-keV X-ray line and the right thermal relic abundance of DM.  We concentrate on the simplest case where, among the SM matter fields, an SU(2)$_L$-singlet charged lepton couples with the DMs and the charged scalars.
Extensions to cases where other SM matter fields couple with the DMs and the charged scalars are straightforward.

This paper is organized as follows.  In section~\ref{sec:model}, we describe our DM model by introducing the new particles, a pair of DM particles and a pair of charged scalars, as well as their interactions with one SM lepton, from which we derive the required magnetic dipole operator for the 3.5-keV X-ray line from DM decays.  In section~\ref{sec:relic}, we calculate the thermal relic density of the DM particles, from which correlation between the new interaction strength and the DM mass is obtained.  Section~\ref{sec:decay} concerns with the decay rate of the heavier DM particle to the lighter one and the 3.5-keV photon, which demands CP-violating couplings in the new interactions.  Combining the current data of the DM relic abundance and the 3.5-keV X-ray line, we perform a numerical analysis to find viable parameter regions in our model.  In section~\ref{sec:pheno}, we evaluate the electric dipole moment of the SM lepton that couples with the DM particles, and perform a collider simulation for the signature of charged scalar pair production at the 14-TeV LHC.  Section~\ref{sec:summary} summarizes our findings.

\section{The Model \label{sec:model}}

\ \ \ We introduce two spin-1/2 Majorana fields $\chi_1, \chi_2$, which are neutral under the SM gauge groups, 
 and two scalar fields $S_1, S_2$ with hypercharge $Y=+1$ but otherwise neutral under SU(3)$_C$ and SU(2)$_L$.
A $Z_2$ symmetry is imposed, under which $\chi_i$, $S_i$ ($i=1,2$) are odd and the SM fields are even.%
\footnote{
This $Z_2$ symmetry may be the remnant of a new U(1)$_X$ gauge symmetry under which $\chi_i$, $S_i$ are oddly charged while all the SM fields are evenly charged.
}
$\chi_i$ and $S_i$ are the mass eigenstates with eigenvalues $m_i$ and $M_i$ ($i=1,2$), respectively.
The masses $m_1,m_2$ and $M_1,M_2$ are of $O(100)$~GeV, with $m_1,m_2$ being smaller than $M_1,M_2$.
Both $\chi_1, \chi_2$ are assumed to be stable at the scale of the age of the Universe ($t\simeq4.4\times10^{17}$~s) and constitute the DM of the Universe.
It is further assumed that $m_2$ is larger than $m_1$ by about 3.5~keV, and that $\chi_2$ decays into $\chi_1$ and a 3.5-keV photon at a rate below the inverse of the age of the Universe, giving rise to the observed 3.5-keV monochromatic photon flux.
Hence the masses of $\chi_1$ and $\chi_2$ are highly degenerate with a fine-tuning of order $\sim$ 3.5~keV$/$100~GeV $\sim 10^{-8}$.
We shall leave the issue of naturally explaining this mass degeneracy for future studies, and regard the degeneracy as a working assumption of the model.

The interaction Hamiltonian involving the fields $\chi_i$ and $S_i$ is given by
\begin{align}
{\cal H}_{{\rm int}} \ &= \ \lambda_{ij} \, \bar{\chi}_i \, \ell_R \, S_j \ + \ {\rm h.c.}~,
\qquad  (i,j=1,2)
\label{int}
\end{align}
where $\ell_R$ denotes a SM SU(2)$_L$-singlet charged lepton, and repeated indices are understood to be summed over.
It could be the case that $\chi_i$ and $S_i$ ($i=1,2$) couple with charged leptons of different flavors.
However, this would give rise to lepton flavor-violating processes that are under severe experimental constraints.
To evade such constraints and for simplicity, we assume that $\chi_i$ and $S_i$ ($i=1,2$) couple with only one SM lepton.  
We note that the diagonal couplings $\lambda_{11}$ and $\lambda_{22}$ can be made real and positive by rotating the phases of $S_i$, 
 whereas the off-diagonal couplings $\lambda_{12}$ and $\lambda_{21}$ are generally complex and can potentially lead to CP-violating phenomena for the SM lepton.

The DM particles $\chi_1, \chi_2$ interact with the SM sector through the interactions in Eq.~(\ref{int}).
With appropriate choices of the interaction strengths, the thermal relic abundance of $\chi_1$ and $\chi_2$ can fit the observed DM abundance.

At one-loop level, the following effective interaction is induced:
\begin{align}
{\cal H}_{eff} \ &= \ C \, \bar{\chi}_1 \, \sigma_{\mu \nu} \, \chi_2 \, F^{\mu \nu}~,
\label{eff int}
\end{align}
where the coefficient $C$ can be expressed in terms of the coupling constants $\lambda_{ij}$ and mass eigenvalues $m_i, M_i$ ($i,j=1,2$) when the SM lepton mass is neglected.
The interaction in Eq.~(\ref{eff int}) gives rise to the decay process $\chi_2 \to \chi_1 \ \gamma$ that can explain the observed 3.5-keV X-ray line.  
We note in passing that the transition electric dipole coupling is not induced in our model because only one chirality of the charged lepton is involved.

In our model, the Majorana DM particles contribute to the spin-independent cross section for DM-nucleon elastic scattering 
 only through the transition magnetic dipole coupling Eq.~(\ref{eff int}).
However, the effective coupling of Eq.~(\ref{eff int}) should be extremely small to explain the 3.5-keV X-ray line.
Therefore, the contribution to the spin-independent cross section is suppressed, and we are allowed to neglect the experimental bounds from DM direct detection experiments.
\\

\section{Relic Abundance \label{sec:relic}}

\ \ \ At high temperatures in the early Universe, the DM particles $\chi_1$, $\chi_2$ are in thermal equilibrium with the SM particles through the following processes with $t$-channel exchanges of the charged scalars $S_1,S_2$:
\begin{align}
\chi_1 \chi_1 \ &\leftrightarrow \ \ell \bar{\ell} ~, \ \ \ 
\chi_2 \chi_2 \ \leftrightarrow \ \ell \bar{\ell} ~.
\label{annihilation}
\end{align}
Processes involving both $\chi_1$ and $\chi_2$ are negligibly rare, because the coupling constants $\lambda_{12}$ and $\lambda_{21}$ must be extremely suppressed in order to explain the 3.5-keV X-ray line observed today.
Below the temperature of $T_{dec} \sim m_1/20 \simeq m_2/20$, the DM particles decouple from the thermal bath and their abundance freezes out.

The squared amplitude for the process $\chi_i \chi_i \leftrightarrow \ell \bar{\ell}$ ($i=1,2$) with spin summations over \textit{initial} and final state particles is given by \cite{oy}
\begin{align}
\vert {\cal M}_i(s) \vert^2 \ &= \ \frac{\vert \lambda_{ii} \vert^4}{16 \pi} \ 4 \
\left[ (M_i^2 - m_i^2)^2 + \frac{s}{2} \ m_i^2 \right] \nonumber \\
&~~\times \ \frac{ \frac{s}2 \sqrt{ 1 - \frac{4m_i^2}{s} } \left( \frac{s}{2} - m_i^2 + M_i^2 \right)
- \left\{ (M_i^2 - m_i^2)^2 + s \, M_i^2 \right\} 
\tan^{-1} \left( \frac{ s/2 \sqrt{1 - 4m_i^2 / s} }{ s/2 - m_i^2 + M_i^2 } \right)
\ }
{ \ \frac{s}{4} \sqrt{ 1 - \frac{4m_i^2}{s} } \ \left( \frac{s}{2} - m_i^2 + M_i^2 \right) \ 
\left\{ (M_i^2 - m_i^2)^2 + s \, M_i^2 \right\} \ } ~, \nonumber \\
\label{amplitude}
\end{align}
where $s = (p_1 + p_2)^2$ with $p_{1,2}$ being the 4-momenta of the initial-state particles.
Assuming that the SM lepton $\ell$ remains in thermal equilibrium and neglecting its mass, the thermally-averaged rate of the pair-annihilation process $\chi_i \chi_i \rightarrow \ell \bar{\ell}$ ($i=1,2$) per pair of $\chi_i$ particles at temperature $T$ is given by~\cite{Edsjo:1997bg}
\begin{align}
\langle \sigma v_i \rangle_T
\ &= \ \frac{1}{ ( n^{{\rm eq}}_{\chi_i}(z))^2 } \ \frac{T}{32 \pi^4} \ \int^{\infty}_{4 m_i^2} {\rm d}s \ 
\sqrt{ \frac{s}{4} - m_i^2 } \ K_1 \left( \frac{\sqrt{s}}{T} \right) \ \vert {\cal M}_i(s) \vert^2 \ ,
\label{sv}
\end{align}
where $n_{\chi_i}^{{\rm eq}}$ is the density of the DM particle $\chi_i$ in thermal equilibrium, 
\begin{align}
n_{\chi_i}^{{\rm eq}} \ &= \ g_{\chi} \int \frac{{\rm d}^3\vec{p}}{(2\pi)^3} \ e^{-\sqrt{\vert \vec{p} \vert^2+m_i^2}/T}
\end{align}
with $g_{\chi} = 2$ being the number of degrees of freedom for $\chi_1$ or $\chi_2$, and $K_1(x)$ is the modified Bessel function of the second kind of order 1.

We apply an approximation formula given in Ref.~\cite{kolb turner} to estimate the relic abundance of $\chi_1$ and $\chi_2$.
To use the formula, we note that $\langle \sigma v_i \rangle_T$ can be approximated for $m_i/T \gtrsim 3$ as 
\begin{align}
\langle \sigma v_i \rangle_T \ &\simeq \ \vert \lambda_{ii} \vert^4 \, \sigma_{i \, 0}(M_i) \, \left(\frac{m_i}{T}\right)^{-3},
\end{align}
 where $\sigma_{i \, 0}$ depends only on $M_i$.
Then the mass density of the DM particle $\chi_i$ ($i=1,2$) is given by
\begin{align}
\Omega_{\chi_i} h^2 \ &\simeq \ 1.07 \times 10^9 \frac{(n+1)x_f^{n+1}}{(g_{*S}/\sqrt{g_*})M_{Pl} \vert \lambda_{ii} \vert^4 \sigma_{i \, 0}} \, {\rm GeV},
\label{chi density}
\end{align}
 where $x^i_f$ corresponds to an estimated decoupling temperature as $x^i_f = m_i/T_{dec}$ and is given by
\begin{align}
x^i_f \ &\simeq \ \log[ \, 0.038(n+1)(g_{\chi}/\sqrt{g_*}) M_{Pl} m_i \vert \lambda_{ii} \vert^4 \sigma_{i \, 0} \, ] \nonumber \\
 &\qquad- \ \left( n+\frac{1}{2} \right) \log[ \, \log[ \, 0.038(n+1)(g_{\chi}/\sqrt{g_*}) M_{Pl} m_i \vert \lambda_{ii} \vert^4 \sigma_{i \, 0} \, ] \, ] \ ,
\end{align}
with $n=3$ here, $M_{Pl}$ is the Planck mass, and $g_*, g_{*S} = 86.25$ are respectively the numbers of relativistic degrees of freedom for the energy density and entropy density at about the decoupling temperature $T_{dec} \sim m_i/20 \sim O(10)$~GeV.
Even though $\chi_2$ eventually decays into $\chi_1$, the net mass density of $\chi_1$ and $\chi_2$ is almost invariant throughout the evolution of the Universe as their mass difference is negligibly small compared to their masses.

Recently, the Planck Collaboration \cite{planck} reported the following DM mass density:
\begin{align}
\Omega_{DM} h^2 \ &= \ 0.1199 \pm 0.0027 ~.
\end{align}
For given values of $m_1,m_2$ and $M_1,M_2$, one can obtain information on the values of $\lambda_{11}, \lambda_{22}$ 
 from Eq.~(\ref{chi density}) and the relation
\begin{align}
\Omega_{\chi_1} h^2 + \Omega_{\chi_2} h^2 \ &= \ \Omega_{DM} h^2 ~.
\label{info from abundance}
\end{align}

For example, if the DM mass is given by $m_1 \simeq m_2 = 300$~GeV and the charged scalar mass by $M_1 \simeq M_2 = 400$~GeV,
 then the right thermal relic abundance of DM can be reproduced with the coupling constants of $\lambda_{11} = \lambda_{22} \simeq 0.5$. 
\\

\section{Dark Matter Decay Rate \label{sec:decay}}

\ \ \ The heavier DM particle $\chi_2$ decays into the lighter one $\chi_1$ and the photon through the effective interaction in Eq.~(\ref{eff int}) induced at one-loop level.
By assuming negligible mass difference between $M_1$ and $M_2$ and that between $m_1$ and $m_2$ when allowed, and defining $M \equiv M_1 \simeq M_2$, $m \equiv m_1 \simeq m_2$, we obtain the following decay rate of $\chi_2$:
\begin{align}
\Gamma_{\chi_2} \ &= \ \frac{k^3}{\pi} \left( \frac{1}{32 \pi^2}
\frac{m}{M^2} \frac{x+\log(1-x)}{2x^2} \right)^2 \, Q^2 \,
{\rm Im}(\lambda_{1 1} \lambda_{2 1}^* + \lambda_{1 2} \lambda_{2 2}^*)^2
\label{rate}
\end{align}
where $k$ denotes the energy of the photon and is given by $k=m_2-m_1 = 3.5$~keV.
$Q$ denotes the electric charge of the lepton, $Q=-e$, and $x$ is defined as $x \equiv m^2/M^2$.
\footnote{
$\chi_2$ can also decay into a $\chi_1$ and two neutrinos.  However, since an off-shell $Z$ boson is involved, the partial width is negligibly small compared to that of the $\chi_2 \rightarrow \chi_1 \gamma$ process.
}

Based upon the assumption that the observed 3.5-keV X-ray line originates from the decay of a sterile neutrino, Refs.~\cite{report1, report2} have derived the mixing angle of the sterile neutrino with a SM active neutrino.
Since the ratio of the DM decay rate over the DM mass appearing in the above calculation is a model-independent quantity, we are allowed to exploit that ratio for our model.
According to Ref.~\cite{report1}, the ratio in our model assumes the following central value:
\begin{align}
\frac{\Gamma_{\chi_2}}{m} \ &= \ 2.4 \times 10^{-29} \ {\rm s}^{-1} \, {\rm keV}^{-1} .
\label{sterile}
\end{align}
On applying the result in Eq.~(\ref{sterile}) to our model, we should take into account the decrease in the number of $\chi_2$ due to its decay, as the lifetime of $\chi_2$ may not be much longer than the age of the Universe in general.
Also, we should note that the net mass density of $\chi_1$ and $\chi_2$ does correspond to the observed DM mass density, and that it is essentially invariant throughout the evolution of the Universe in view of their negligible mass difference.
We thus obtain the following relation:
\begin{align}
\frac{\Omega^{ini}_{\chi_2}h^2 \ \exp(-\Gamma_{\chi_2} t_{cos})}{\Omega^{ini}_{\chi_1}h^2 + \Omega^{ini}_{\chi_2}h^2} \ \frac{\Gamma_{\chi_2}}{m} \ &= \ 2.4 \times 10^{-29} \ {\rm s}^{-1} \, {\rm keV}^{-1} .
\label{flux}
\end{align}
 where $t_{cos} \simeq 4.4\times10^{17}$~s, and $\Omega^{ini}_{\chi_i}h^2$ denotes the mass density of $\chi_i$ ($i=1,2$) when it freezes out.
If $\chi_1$ and $\chi_2$ decouple from the thermal bath in the early Universe in the same way and the relation $\Omega^{ini}_{\chi_1}h^2=\Omega^{ini}_{\chi_2}h^2$ is satisfied, Eq.~(\ref{flux}) 
 can be reduced to
\begin{align}
\Gamma_{\chi_2} t_{cos} \exp(-\Gamma_{\chi_2} t_{cos})
= \left( \frac{m_2}{100 \, {\rm GeV}} \right) \times 2.1 \times 10^{-4}~,
\end{align}
 which has two solutions of $\Gamma_{\chi_2}$ for each value of $m_2$.  Considering the range of 100~GeV $\le m_2 \le$ 1000~GeV, one solution is
\begin{align}
\Gamma_{\chi_2} \ &\simeq \ \frac{m_2}{100 \, {\rm GeV}} \times 4.8 \times 10^{-21} \ {\rm s}^{-1} ~, 
\label{rate solution1} 
\end{align}
and the other is numerically evaluated as
\begin{align}
\Gamma_{\chi_2} \ &\simeq \ 1.9 \times 10^{-17} \ {\rm s}^{-1}
~\mbox{ to }~ 1.3 \times 10^{-17} \ {\rm s}^{-1}
\label{rate solution2}
\end{align}
as $m_2$ varies from 100~GeV and 1~TeV.
For the solution Eq.~(\ref{rate solution2}), the value of $\Gamma_{\chi_2}$ lies between $1.3 \times 10^{-17}$~s$^{-1}$ and $1.9 \times 10^{-17}$~s$^{-1}$ for $m_2$ between 1000~GeV and 100~GeV. 
The solution Eq.~(\ref{rate solution1}) corresponds to the case where the diminution of $\chi_2$ up to the present is negligible, and the decay rate of $\chi_2$ controls the flux of the 3.5-keV X-ray line.
On the other hand, the solution Eq.~(\ref{rate solution2}) implies that the number of $\chi_2$ has decreased in such a way that its number density at present gives the right flux of the 3.5-keV X-ray line.
To estimate how these solutions depend on the age of the Universe, $t_{cos}$, we differentiate both sides of Eq.~(\ref{flux}) by $t_{cos}$ while fixing the mass of $\chi_2$ and the observed flux of the 3.5-keV X-ray,
 and derive the quantity $\Delta \, \equiv \, (t_{cos}/\Gamma_{\chi_2}) \vert \partial \Gamma_{\chi_2}/ \partial t_{cos} \vert$, a measure of fine-tuning.
If $\Delta$ is $O(1)$ or above, this means that the solution for $\Gamma_{\chi_2}$ is connected with the age of the Universe, which is a miraculous coincidence that we do not expect to occur,
 and hence the solution is discarded.
We have
\begin{align}
\Delta \ &= \ \frac{t_{cos}}{\Gamma_{\chi_2}} \left\vert \frac{\partial \Gamma_{\chi_2}}{\partial t_{cos}} \right\vert
\ = \ \frac{t_{cos} \Gamma_{\chi_2}}{\vert 1 - t_{cos}\Gamma_{\chi_2} \vert}.
\end{align}
For the solution Eq.~(\ref{rate solution1}), $\Delta \lesssim 10^{-2}$, 
 whereas for the solution Eq.~(\ref{rate solution2}), $\Delta$ is $O(1)$.
We thus dismiss the solution Eq.~(\ref{rate solution2}) for being too closely related to the age of the Universe,
 and only adopt the solution Eq.~(\ref{rate solution1}).
\\

\section{Numerical Analysis on the Coupling Constants \label{sec:coupling}}

\begin{figure}[thbp]
 \begin{minipage}{0.5\hsize}
  \begin{center}
   \includegraphics[width=75mm]{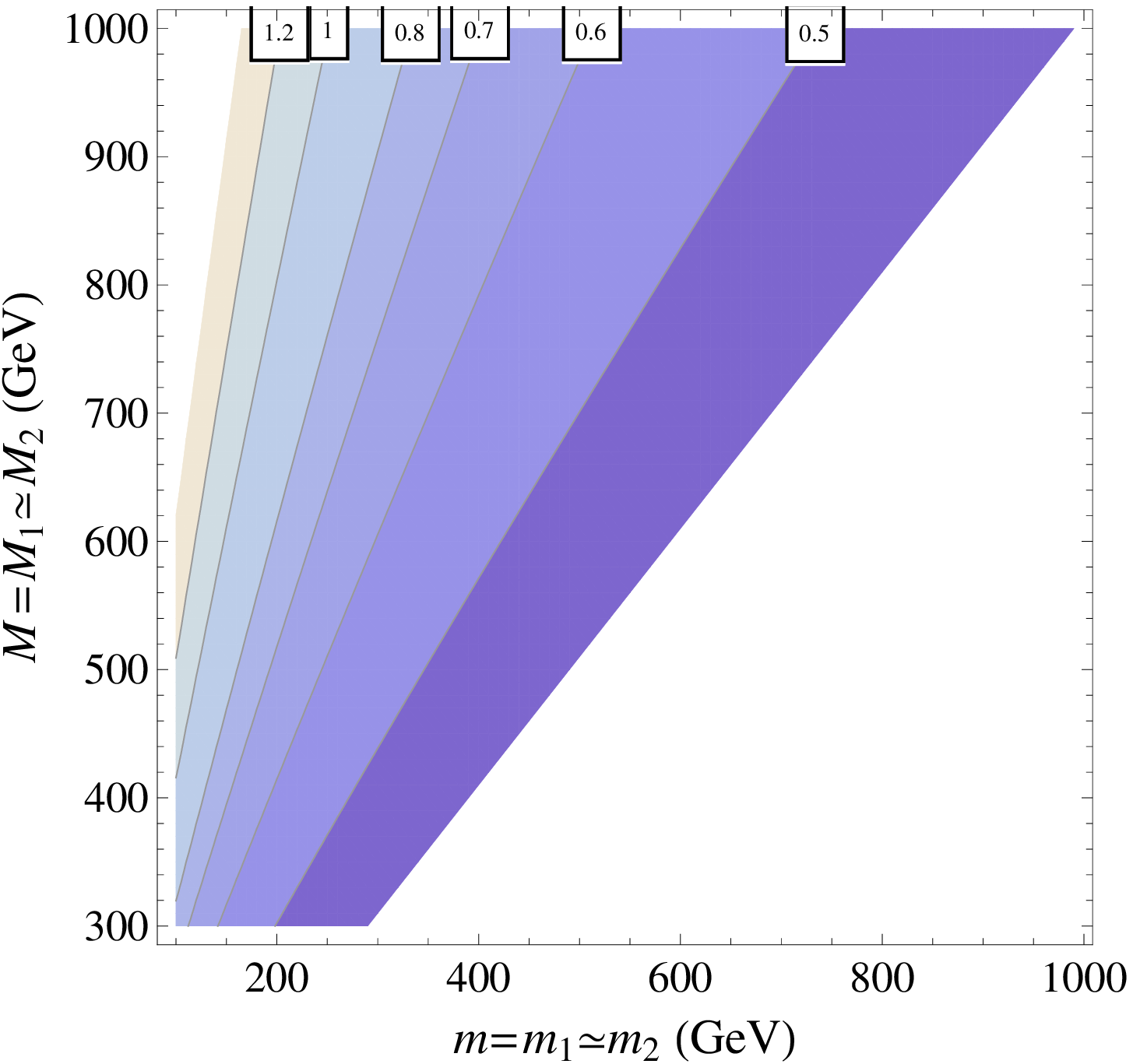}
  \end{center}
 \end{minipage}
  \begin{minipage}{0.5\hsize}
  \begin{center}
   \includegraphics[width=75mm]{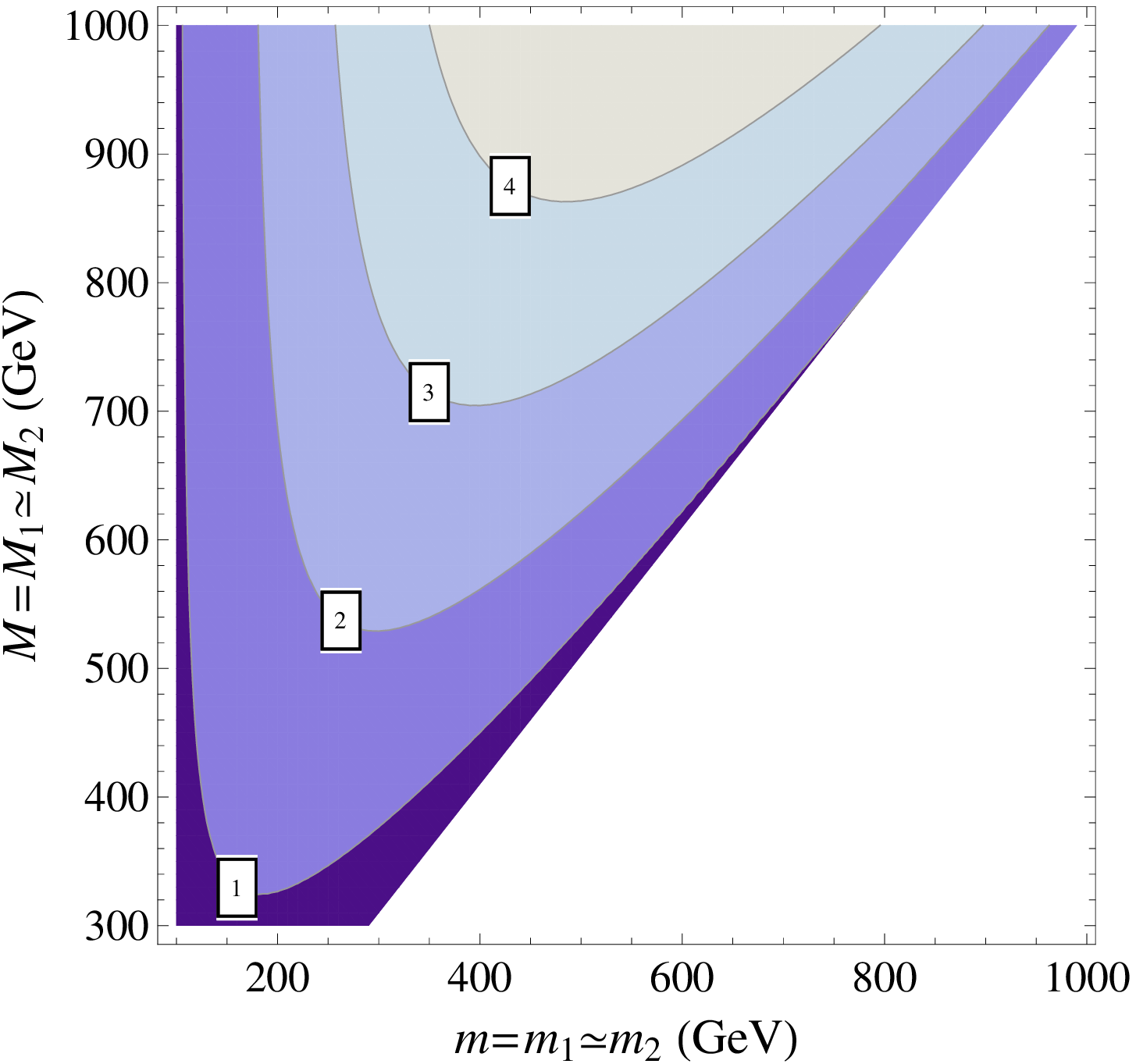}
  \end{center}
 \end{minipage}
 \caption{
  [Left] Contour plot of the diagonal coupling constants $\lambda=\lambda_{1 1}=\lambda_{2 2}$ that
   give the net thermal relic abundance of DM particles $\chi_1$ and $\chi_2$ that fits the Planck data,
   on the plane of the DM particle mass $m=m_1\simeq m_2$ and the charged scalar mass $M=M_1 \simeq M_2$.
   In the upper left blank corner of the parameter space, the right DM thermal relic abundance cannot be obtained.
  [Right] Contour plot of $\epsilon$ in units of $10^{-8}$, on the $m$-$M$ plane.  The off-diagonal coupling constants reproduce the correct ratio of the DM decay rate over the DM mass inferred from the flux of the 3.5-keV X-ray emission line from galaxy clusters.
  \label{lam-ep}
  }
\end{figure}

\ \ \ From Eqs.~(\ref{chi density}), (\ref{info from abundance}), (\ref{rate}) and (\ref{flux}), we can obtain information on the coupling constants $\lambda_{ij}$ ($i,j=1,2$) for a given set of values of $m_1, m_2, M_1, M_2$.  In this section, we perform a numerical analysis on $\lambda_{ij}$ to confirm that our model can explain both the 3.5-keV X-ray line and the DM abundance with specific values of the coupling constants.  To simplify the analysis, we assume $\lambda_{1 1}=\lambda_{2 2}\equiv \lambda$ and $\lambda_{1 2}=-\lambda_{2 1}\equiv i \epsilon \lambda$, where $\lambda > 0$ and $\epsilon$ is a small real parameter.
Also, we keep using the approximations that $M_1=M_2 \equiv M$ and that $m_1=m_2 \equiv m$ when we are allowed to neglect $m_2-m_1$.
With these simplifications, we are able to derive $\lambda$ and $\epsilon$ for each set of values for $m$ and $M$.
In this study, we concentrate on the parameter region where 300~GeV~$<M<$~1~TeV and $100$~GeV~$<m<M-10$~GeV.

The left plot in figure~\ref{lam-ep} is a contour plot of the diagonal coupling constants $\lambda$ on the $m$-$M$ plane, derived by requiring that the net thermal relic abundance of $\chi_1$ and $\chi_2$ fit the observed DM abundance $\Omega_{DM} h^2 = 0.1199$.
The right plot in figure~\ref{lam-ep} is a contour plot of $\epsilon$ on the $m$-$M$ plane, derived by requiring that the photon emission in the decay of $\chi_2$ reproduce the observed flux of the 3.5-keV X-ray.
The numbers on the contours in the right plot are displayed in units of $10^{-8}$.  Therefore, the off-diagonal couplings are typically smaller in size than the diagonal ones by eight orders of magnitude.
%
%

\section{Phenomenology \label{sec:pheno}}

\ \ \ In this section, we discuss experimental signatures of the model other than the 3.5-keV emission line.

\subsection{Electric Dipole Moments \label{sec:EDM}}

\ \ \ We observe from Eq.~(\ref{rate}) that the decay rate of $\chi_2$ is proportional to the square of the imaginary part of $\lambda_{1 1} \lambda_{2 1}^* + \lambda_{1 2} \lambda_{2 2}^*$.  By rotating the phases of $S_1, S_2$, we can always take $\lambda_{11}, \lambda_{22}$ to be real.  Then the decay rate is non-zero only when $\lambda_{12} \neq \lambda_{12}^*$ or $\lambda_{21} \neq \lambda_{21}^*$ is satisfied; that is, only when the interaction in Eq.~(\ref{int}) violates CP symmetry.  This is an important consequence of the Majorana nature of our DM particles: the transition magnetic dipole interaction of the Majorana DM particles Eq.~(\ref{eff int}), $\bar{\chi}_1 \sigma_{\mu \nu} \chi_2 \, F^{\mu \nu}$, which is induced at the one-loop level, flips sign under the CP transformation.  Hence its effective coupling constant should be proportional to the amount of CP violation.

The CP-violating parts in Eq.~(\ref{int}) contribute to the electric dipole moment (EDM) of the SM lepton involved in the interaction.  This occurs at the two-loop level.  More importantly, this contribution is proportional to the mass difference between the two DM particles, $m_2-m_1 \simeq 3.5$~keV, which can be seen as follows.  Each of the two-loop diagrams contributing to the lepton EDM is proportional to either $\lambda_{11} \lambda_{21}^* \lambda_{12}^* \lambda_{22}$ or $\lambda_{12} \lambda_{22}^* \lambda_{11}^* \lambda_{21}$.  By exchanging the roles of $\chi_1$ and $\chi_2$ in the loop of each diagram, a diagram proportional to $\lambda_{11} \lambda_{21}^* \lambda_{12}^* \lambda_{22}$ becomes one proportional to $\lambda_{12} \lambda_{22}^* \lambda_{11}^* \lambda_{21}$, and vice versa.  
Therefore, if the masses of $\chi_1$ and $\chi_2$ were the same, the sum of the two-loop lepton EDM diagrams would be proportional to $\lambda_{11} \lambda_{21}^* \lambda_{12}^* \lambda_{22} + \lambda_{12} \lambda_{22}^* \lambda_{11}^* \lambda_{21}$, which is a quantity invariant under the CP transformation, implying that the amplitude for the lepton EDM would vanish.  Hence the amplitude for the lepton EDM should actually be proportional to the mass difference between $\chi_1$ and $\chi_2$.  In fact, the same argument applies to all orders in perturbation theory, and we conclude that the leading contribution to the lepton EDM arises at the two-loop level and is proportional to $m_2-m_1$.

We can estimate the magnitude of the lepton EDM arising from the interaction of Eq.~(\ref{int}) by a dimensional analysis.  Assuming that the charged scalar masses and the DM masses are of the same order of magnitude, denoted by $M$, we have 
\begin{align}
d_{EDM} \ &\sim \ e \ \left( \frac{1}{16\pi^2} \right)^2 \, \frac{1}{M^2} \, (m_2 - m_1) \, {\rm Im}(\lambda_{11} \lambda_{21}^* \lambda_{12}^* \lambda_{22})~.
\label{edm}
\end{align}
To make an aggressive estimate, we take $\lambda_{11}=\lambda_{22}=1$ and Im$(\lambda_{12})=-$Im $(\lambda_{21})=5\times10^{-8}$ as inferred from figure~\ref{lam-ep}, and assume Re$(\lambda_{12})=-$Re$(\lambda_{21})=1$.  Then, substituting $m_2-m_1=3.5$~keV, we have
\begin{align}
d_{EDM} \ &\sim \ 3 \times 10^{-35} \, \left( \frac{100 \, {\rm GeV}}{M} \right)^2 \ \mbox{$e$-cm} \ .
\end{align}
Even when the DMs and the charged scalars couple with the SM electron, i.e., $\ell_R$ appearing in Eq.~(\ref{int}) is the electron, the EDM is so small for $M \simeq 100$~GeV that our model safely evades the current experimental bound.

\subsection{Signatures at the LHC \label{sec:LHC}}

\ \ \ In collider experiments, the most distinctive signature of our model is the existence of two charged scalars with the mass of $O(100)$~GeV that couple with DM particles and the \textit{same} SM lepton.  The most prominent process is the s-channel pair production of the charged scalars, followed by their decays into DM particles and leptons.  Therefore, the signature would be a pair of opposite-sign leptons plus missing energy.  In this subsection, we concentrate on the case where the charged scalars and the DMs couple with the muon for the ease of being identified particularly at hadron colliders.  In our following simulations, we consider the 14-TeV LHC.

There are two strategies for testing the model at the LHC, depending on the mass degeneracy of the two charged scalars.  (Although we assume in Section~\ref{sec:coupling} that the charged scalar masses are equal, this is not mandatory in the model.)  
If the masses of the two charged scalars are different, one can confirm the existence of two charged scalar particles by reconstructing their masses using the $M_{T2}$ variable~\cite{mt2}.  On the other hand, even if they are degenerate, one can still gain a hint of the mass degeneracy by measuring the mass, calculating the cross section for a charged scalar pair of that mass in $pp$ collisions, and comparing it with the measured cross section of the signal events, although these require accurate measurements of mass and cross section and precise theoretical calculations.  In this subsection, we focus on the first case, and examine the possibility of testing the model at the 14-TeV LHC by simulating the charged scalar production events as well as SM background events.

Before plunging into the study on the observability of the model's signature at the LHC, we first discuss the current bounds on the model from the 8-TeV LHC experiments.  The above-mentioned signature process of the charged scalar production has an event topology similar to the direct slepton (supersymmetric partner of the SM lepton) pair production followed by their prompt decays into stable neutralinos and a pair of SM leptons with opposite signs, as has been searched for by the ATLAS Collaboration \cite{atlas} and the CMS Collaboration \cite{cms}.

The result reported by the ATLAS Collaboration was based upon the 8-TeV data with an integrated luminosity of 20.3~fb$^{-1}$, and focused on the signal regions termed `SR-$m_{{\rm T}2}$'.  The CMS Collaboration, on the other hand, used 19.5~fb$^{-1}$ of data for their search analysis.  Plot~(a) of Fig.~8 in Ref.~\cite{atlas} and the bottom plot of Fig.~18 in Ref.~\cite{cms} respectively give a lower bound of about 250~GeV and 190~GeV on the slepton mass at 95\% confidence level.

Because of the existence of the \textit{two} charged scalars and if the scalar masses are the same, the signal cross section in our model will double that of a supersymmetric model with only \textit{one} flavor of SU(2)$_L$-singlet slepton decaying into the neutralino and the lepton.
On the other hand, in our leading-order calculation with MadGraph5~\cite{mg5}, the production cross section of the slepton pair in $pp$ collisions with $\sqrt{s}=8$~TeV decreases by a factor of 2 as the slepton mass increases from 250~GeV to 300~GeV.
In view of these, our model is seen to safely evade the current bounds from the LHC experiments if the masses of both charged scalars are above 300~GeV.

In the numerical study, we use MadGraph5~\cite{mg5} for parton-level event generations, PYTHIA8~\cite{pythia} for simulating parton showering and hadronization, and PGS4~\cite{pgs4} for detector simulations.  We consider the following two benchmark mass spectra and coupling constants:
\begin{align}
M_1 &= 400~{\rm GeV} ~,~ M_2 = 300~{\rm GeV} ~,~ m_1 \simeq m_2 = 100~{\rm GeV} ~,~ \lambda_{11} = \lambda_{22} = 0.8 ~; \label{bench1} \\
M_1 &= 500~{\rm GeV} ~,~ M_2 = 300~{\rm GeV} ~,~ m_1 \simeq m_2 = 100~{\rm GeV} ~,~ \lambda_{11} = \lambda_{22} = 1.0 ~. \label{bench2}
\end{align}
We generate parton-level events for the following signal processes for $i=1,2$, where $j$ denotes any parton:
\begin{align}
& p \ p \ \rightarrow \ S_i \ S_i^{\dagger}, \ \ \ S_i \ \rightarrow \ \chi_i \ \mu^+, \ \ \ S_i^{\dagger} \ \rightarrow \ \chi_i \ \mu^-,
\nonumber \\
& p \ p \ \rightarrow \ S_i \ S_i^{\dagger} \ j, \ \ \ S_i \ \rightarrow \ \chi_i \ \mu^+, \ \ \ S_i^{\dagger} \ \rightarrow \ \chi_i \ \mu^-,
\ \ \ (i=1,2).
\end{align}
Values of the coupling constants affect collider signatures by changing the widths of the charged scalars, yet such effects are negligible.  The 0-jet and 1-jet events are matched after parton showering by the MLM matching scheme~\cite{mlm}, and the showered and hadronized events are processed for the detector simulation. 
Jets are reconstructed by the anti-$k_T$ jet clustering algorithm \cite{antikt} with $\Delta R=0.4$.
We include both the 0-jet and 1-jet events to take into account the effect of an initial state radiation jet(s),
 the mismeasurement of which causes the uncertainty of missing transverse momentum.

The dominant sources of SM background are diboson $W^+ W^-$, $ZZ$ and $Z\gamma$ production processes.
We therefore generate parton-level events for the following background processes, where $j$ denotes any parton,
\begin{align}
&
p \ p \ \rightarrow \ W^+ \ W^-, \ \ \ W^+ \ \rightarrow \ \mu^+ \nu_{\mu}, \ \ \ W^- \ \rightarrow \ \mu^- \bar{\nu}_{\mu} ~,
\nonumber \\
&
p \ p \ \rightarrow \ W^+ \ W^- \ j, \ \ \ W^+ \ \rightarrow \ \mu^+ \nu_{\mu}, \ \ \ W^- \ \rightarrow \ \mu^- \bar{\nu}_{\mu} ~,
\nonumber \\
&
p \ p \ \rightarrow \ Z \ Z^* / \gamma (\rightarrow \ \mu^+ \mu^-), \ \ \ Z \ \rightarrow \ \nu \bar{\nu} ~,
\nonumber \\
&
p \ p \ \rightarrow \ Z \ Z^* / \gamma (\rightarrow \ \mu^+ \mu^-) \ j, \ \ \ Z \ \rightarrow \ \nu \bar{\nu} ~,
\end{align}
and the generated events are processed in the same way as the signal events.  Since we will impose a selection cut of $m(\mu^+, \, \mu^-) > 150$~GeV, we only generate those events where a $\mu^+ \mu^-$ pair is produced through an off-shell $Z$ boson or a photon.

We impose the following kinetic cuts on the events after detector simulation:
\smallskip
\\
\indent (a) There should be two opposite-sign muons whose pseudo-rapidity and transverse momentum satisfy $\vert \eta_{\mu} \vert < 2.4$ and $p^\mu_T > 25$~GeV.  \\
\indent (b) The missing transvserse momentum should satisfy $\mpt_T > 200$~GeV. \\
\indent (c) The invariant mass of the muons should satisfy $m(\mu^+, \, \mu^-) > 150$~GeV.
\indent (d) The muons should be separated from any jet by the distance of $\Delta R(\vec{p}_{\mu}, \, \vec{p}_j)~>~0.4$.
\smallskip

\begin{figure}[thbp]
 \begin{center}
   \includegraphics[width=75mm]{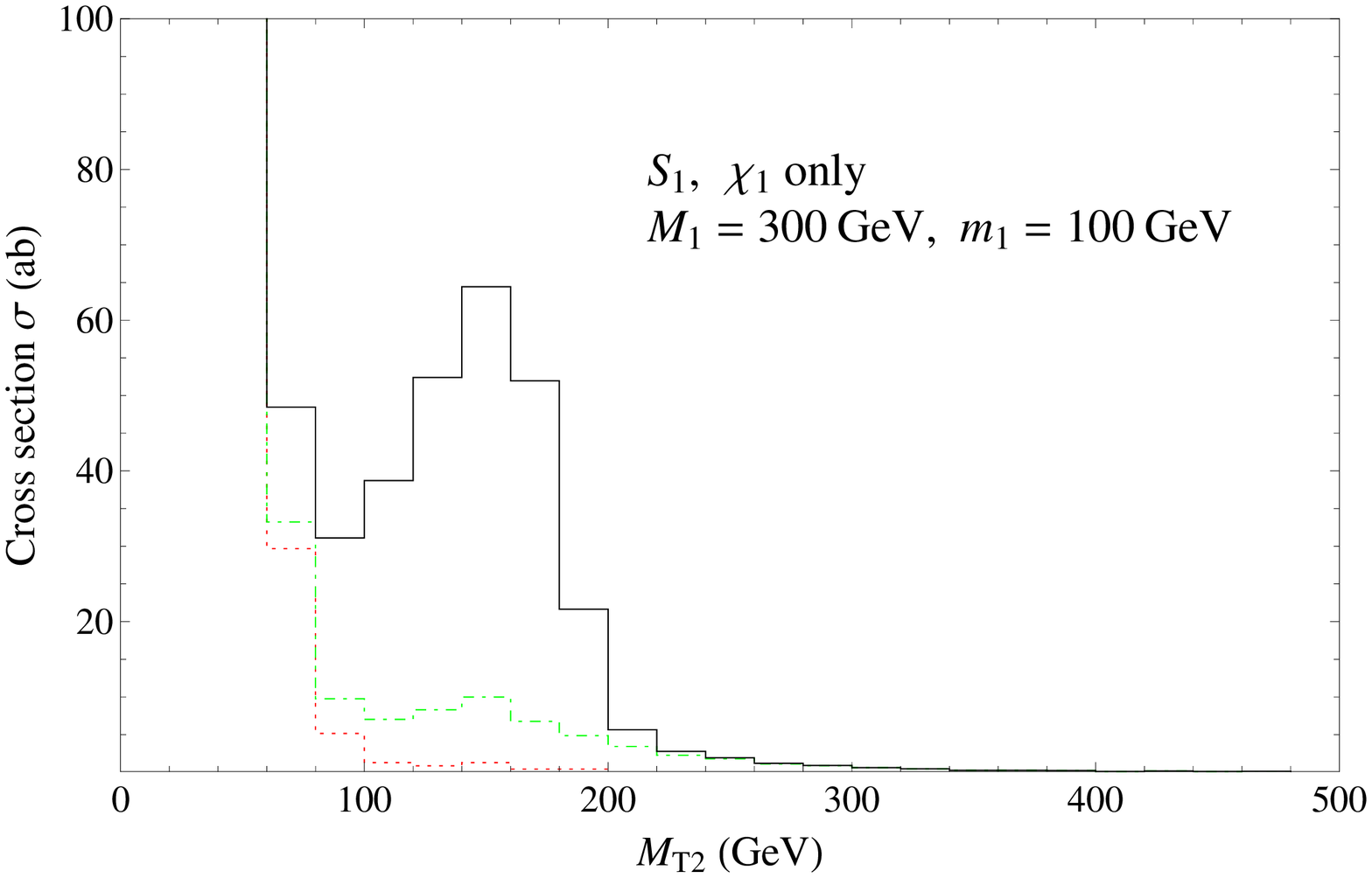}
 \end{center}
 \begin{minipage}{0.5\hsize}
  \begin{center}
   \includegraphics[width=75mm]{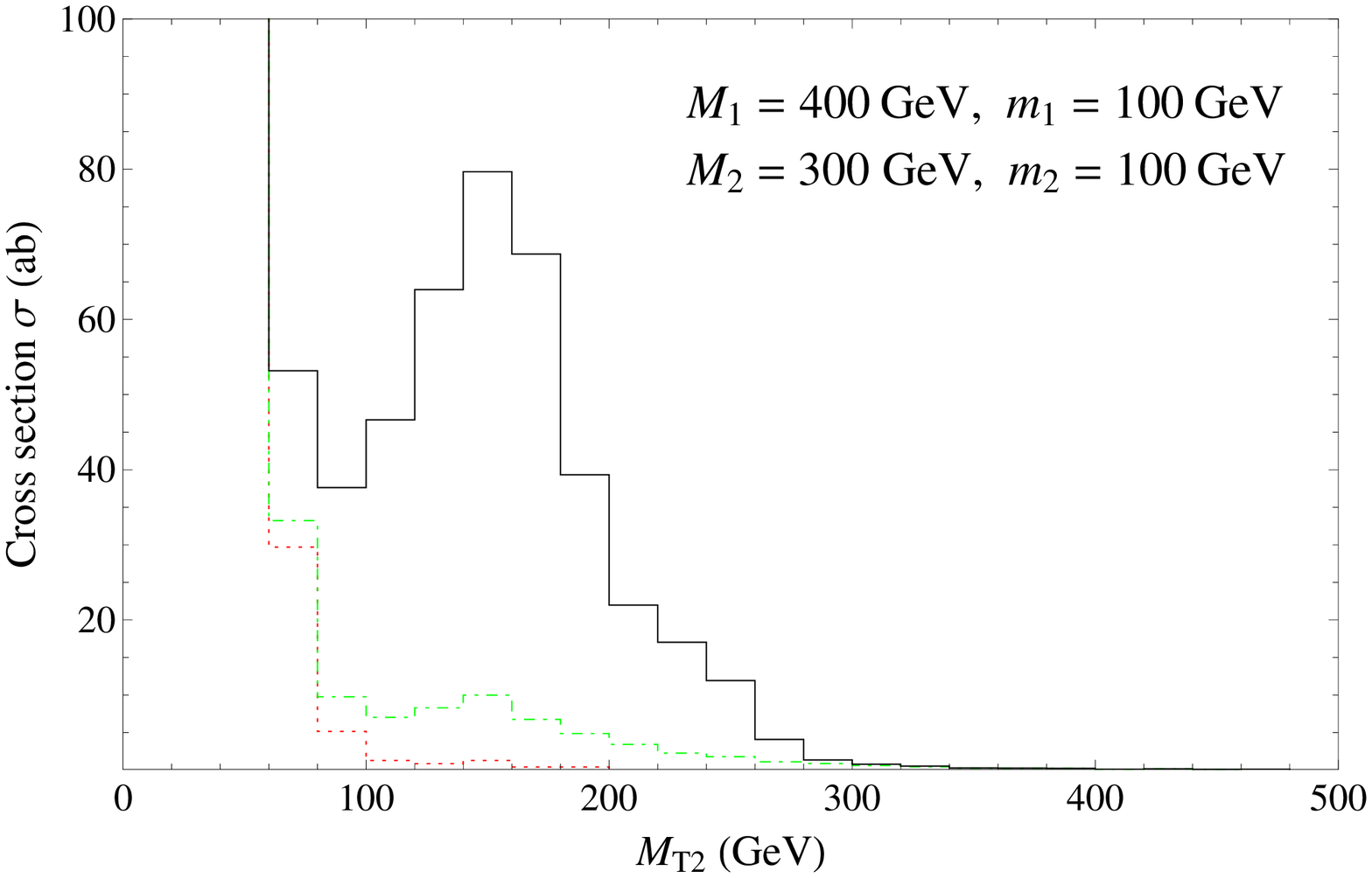}
  \end{center}
 \end{minipage}
  \begin{minipage}{0.5\hsize}
  \begin{center}
   \includegraphics[width=75mm]{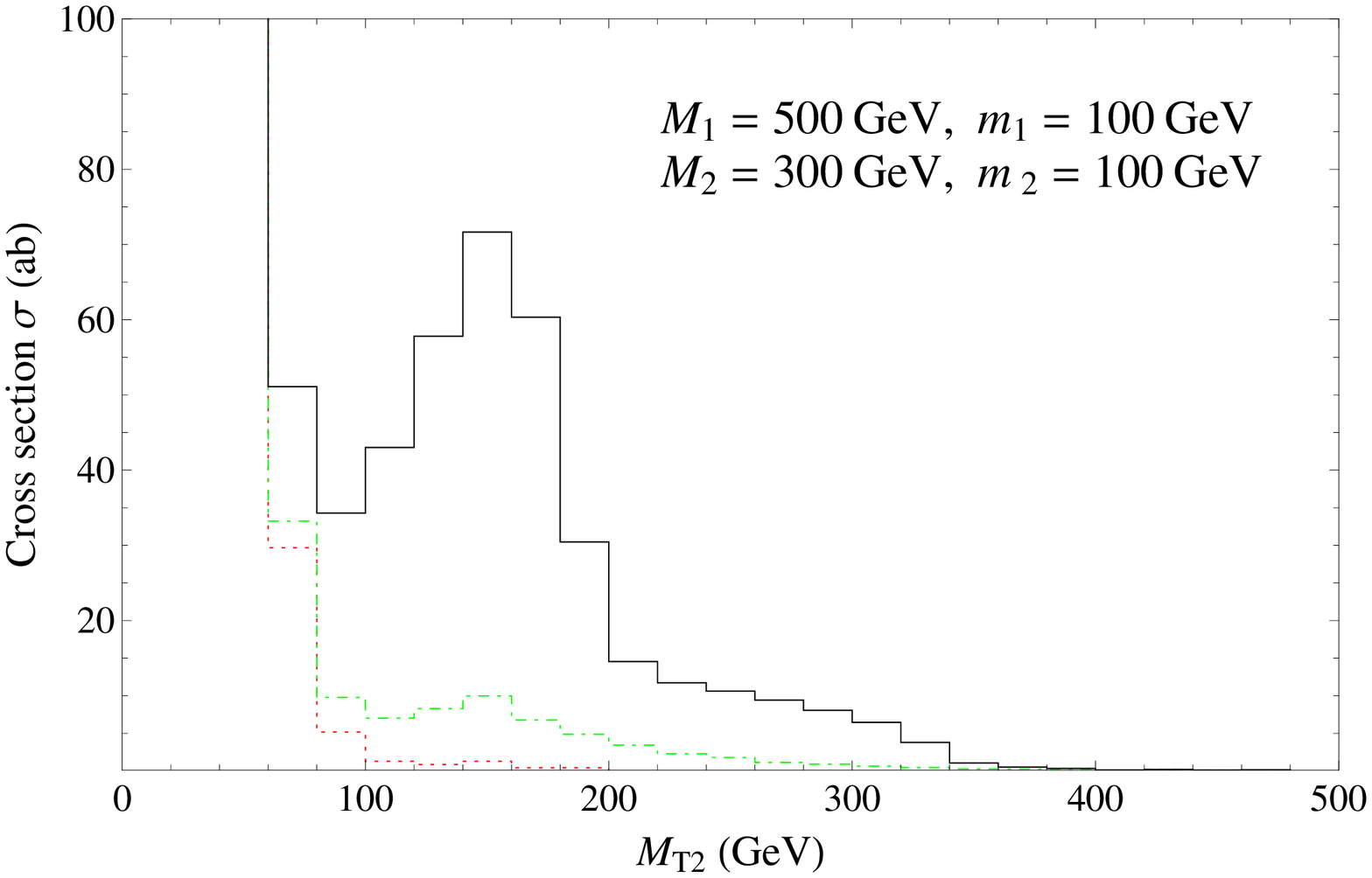}
  \end{center}
 \end{minipage}
 \caption{
 Each plot shows the $M_{T2}$ distribution for the SM $W^+ W^-$ production process (red dotted line), that plus the SM $Z Z/\gamma$ production processes (green dot-dashed line), and those plus the signal $S_i S_i^{\dagger}$ ($i=1,2$) production processes followed by the $S_i$ decay into a muon and a DM particle $\chi_i$, at the 14-TeV LHC.  The simulation is done at the detector level, after imposing the selection cuts (a), (b), and (c) defined in the text.  The upper plot corresponds to the case where there exit only $S_1$ and $\chi_1$ particles whose masses are respectively $M_1=300$~GeV and $m_1=100$~GeV, the lower left plot to the benchmark mass spectrum in Eq.~(\ref{bench1}), and the lower right plot to the benchmark mass spectrum in Eq.~(\ref{bench2}).  The vertical bars in the histograms are in units of ab per 20-GeV bin.}
 \label{mt2}
\end{figure}

We evaluate the $M_{T2}$ variable, defined as
\begin{align}
M_{T2}^2 \ &= \ {\rm min}_{\vec{p}_1 + \vec{p}_2 = \vec{\mpt}_T} 
\ [ \ {\rm max}\{ M_T(\vec{p}_{\mu^+}, \, \vec{p}_1), \ M_T(\vec{p}_{\mu^-}, \, \vec{p}_2) \} \ ],
\label{mt2def}
\end{align}
for each event and plot its distribution for both the signal and background processes in figure~\ref{mt2}.

In figure~\ref{mt2}, we observe a difference in the tails of the $M_{T2}$ distributions for the case where there is only one set of charged scalar and DM particle and the cases where there are two charged scalars of different masses each of which couples with a different DM particle.
Compared to the signals of our benchmark mass spectra in Eqs.~(\ref{bench1}, \ref{bench2}), the SM backgrounds are well suppressed with the cuts (a), (b), and (c).
We therefore conclude that, with $\sim 1$~ab$^{-1}$ of data at the 14-TeV LHC, there is a good chance to confirm the existence of two charged scalar-DM pairs, if the charged scalar masses are separated by $O(100)$~GeV and both of them are below about 500~GeV.
\\

\section{Summary \label{sec:summary}}

\ \ \ We have proposed and studied a model where the unidentified 3.5-keV X-ray emission line from galaxy clusters is explained in terms of the decay of a WIMP DM particle into another WIMP DM particle and a photon.  We have argued that the simplest model is the one that contains two spin-1/2 Majorana DM particles with the mass difference of 3.5~keV and coupling with two charged scalars of $O(100)$~GeV mass and a SM matter fermion(s).  
Each DM particle couples dominantly with a distinct charged scalar and the same SM matter fermion through an $O(0.1)$ coupling constant, which is responsible for the thermalization of the DM in the early Universe and realizes the WIMP DM scenario.  On the other hand, each DM particle also couples with the other charged scalar through a tiny coupling constant of order $10^{-8}$, which induces a dipole transition coupling between the two DM species at one-loop level, thereby giving rise to the decay of the heavier DM into the lighter one and the photon.

As prominent phenomenological signatures of the model, we have discussed the contribution of the new particles to the EDM of the SM matter, and the collider signatures of the charged scalars at the 14-TeV LHC.  We emphasize that, since the dipole transition operator of the Majorana fermions violates CP symmetry, one can infer the CP-violating part of the charged scalar-DM-SM matter fermion couplings from the DM decay rate, which is directly connected with the flux of the 3.5-keV emission line.  We have discovered that the induced lepton EDM is so tiny that it safely evades the current experimental bound even when the SM matter is the electron.  Also, we have studied the signatures of the model at the 14-TeV LHC, paying particular attention to whether one can confirm the existence of the two-component charged scalars when their masses are separate.  By observing the shape of the tail in the $M_{T2}$ distribution, it is possible to distinguish the cases where there is only one set of charged scalar and DM particle that couple with the SM muon and where there are two charged scalar-DM sets, provided the mass difference is of the order of 100~GeV and the charged scalar masses are below about 500~GeV.

\vspace{1cm}
{\it Note Added: } While this work was being written up, we noticed a similar work by Geng et.~al.~\cite{Geng:2014zqa}.  However, our model is different from theirs in the number of charged scalars and the associated interactions.  Therefore, we have different collider signatures for the new particles.  In addition, unlike their model, ours does not introduce new sources for lepton flavor violating processes.
Neither do we have the transition electric dipole coupling of two Majorana DM particles at the one-loop level, and only the transition magnetic dipole coupling appears in our model.

\section*{Acknowledgments}

The authors would like to thank K.~Yagyu for participating in discussions during the early stage of this project.
C.-W.~C. thanks the hospitality of the KITP at Santa Barbara where last part of this work was being finished.  
This research was supported in part by the National Science Council of R.O.C. under Grant Nos.~NSC-100-2628-M-008-003-MY4 and NSC-102-2811-M-008-019 and in part by the National Science Foundation under Grant No. NSF PHY11-25915.


\begin{thebibliography}{99}

\bibitem{report1}
  E.~Bulbul, M.~Markevitch, A.~Foster, R.~K.~Smith, M.~Loewenstein and S.~W.~Randall,
  Astrophys.\ J.\  {\bf 789}, 13 (2014)
  [arXiv:1402.2301 [astro-ph.CO]].

\bibitem{report2}
  A.~Boyarsky, O.~Ruchayskiy, D.~Iakubovskyi and J.~Franse,
  arXiv:1402.4119 [astro-ph.CO].

\bibitem{studies}
  H.~Ishida, K.~S.~Jeong and F.~Takahashi,
  Phys.\ Lett.\ B {\bf 732}, 196 (2014)
  [arXiv:1402.5837 [hep-ph]];
  D.~P.~Finkbeiner and N.~Weiner,
  arXiv:1402.6671 [hep-ph];
  T.~Higaki, K.~S.~Jeong and F.~Takahashi,
  Phys.\ Lett.\ B {\bf 733}, 25 (2014)
  [arXiv:1402.6965 [hep-ph]];
  J.~Jaeckel, J.~Redondo and A.~Ringwald,
  Phys.\ Rev.\ D {\bf 89}, 103511 (2014)
  [arXiv:1402.7335 [hep-ph]];
  H.~M.~Lee, S.~C.~Park and W.~-I.~Park,
  arXiv:1403.0865 [astro-ph.CO];
    K.~N.~Abazajian,
  Phys.\ Rev.\ Lett.\  {\bf 112}, 161303 (2014)
  [arXiv:1403.0954 [astro-ph.CO]];
  C.~El Aisati, T.~Hambye and T.~Scarna,
  arXiv:1403.1280 [hep-ph];
    K.~Hamaguchi, M.~Ibe, T.~T.~Yanagida and N.~Yokozaki,
  arXiv:1403.1398 [hep-ph];
      J.~-C.~Park, S.~C.~Park and K.~Kong,
  Phys.\ Lett.\ B {\bf 733}, 217 (2014)
  [arXiv:1403.1536 [hep-ph]];
    M.~T.~Frandsen, F.~Sannino, I.~M.~Shoemaker and O.~Svendsen,
  JCAP {\bf 1405}, 033 (2014)
  [arXiv:1403.1570 [hep-ph]];
  S.~Baek and H.~Okada,
  arXiv:1403.1710 [hep-ph];
  K.~Nakayama, F.~Takahashi and T.~T.~Yanagida,
  arXiv:1403.1733 [hep-ph];
  K.~-Y.~Choi and O.~Seto,
  Phys.\ Lett.\ B {\bf 735}, 92 (2014)
  [arXiv:1403.1782 [hep-ph]];
    M.~Cicoli, J.~P.~Conlon, M.~C.~D.~Marsh and M.~Rummel,
  arXiv:1403.2370 [hep-ph];
    F.~Bezrukov and D.~Gorbunov,
  arXiv:1403.4638 [hep-ph];
    C.~Kolda and J.~Unwin,
  arXiv:1403.5580 [hep-ph];
  R.~Allahverdi, B.~Dutta and Y.~Gao,
  arXiv:1403.5717 [hep-ph];
  N.~-E.~Bomark and L.~Roszkowski,
  arXiv:1403.6503 [hep-ph];
    S.~P.~Liew,
  JCAP {\bf 1405}, 044 (2014)
  arXiv:1403.6621 [hep-ph];
    Z.~Kang, P.~Ko, T.~Li and Y.~Liu,
  arXiv:1403.7742 [hep-ph];
  S.~V.~Demidov and D.~S.~Gorbunov,
  arXiv:1404.1339 [hep-ph];
    F.~S.~Queiroz and K.~Sinha,
  Phys.\ Lett.\ B {\bf 735}, 69 (2014)
  [arXiv:1404.1400 [hep-ph]];
  E.~Dudas, L.~Heurtier and Y.~Mambrini,
  arXiv:1404.1927 [hep-ph];
  K.~S.~Babu and R.~N.~Mohapatra,
  Phys.\ Rev.\ D {\bf 89}, 115011 (2014)
  [arXiv:1404.2220 [hep-ph]];
  K.~P.~Modak,
  arXiv:1404.3676 [hep-ph];
  J.~M.~Cline, Y.~Farzan, Z.~Liu, G.~D.~Moore and W.~Xue,
  arXiv:1404.3729 [hep-ph];
  H.~Okada and T.~Toma,
  arXiv:1404.4795 [hep-ph];
  H.~M.~Lee,
  arXiv:1404.5446 [hep-ph];
    D.~J.~Robinson and Y.~Tsai,
  arXiv:1404.7118 [hep-ph];
    J.~P.~Conlon and F.~V.~Day,
  arXiv:1404.7741 [hep-ph];
  S.~Baek, P.~Ko and W.~-I.~Park,
  arXiv:1405.3730 [hep-ph];
  K.~Nakayama, F.~Takahashi and T.~T.~Yanagida,
  arXiv:1405.4670 [hep-ph];
  S.~Chakraborty, D.~K.~Ghosh and S.~Roy,
  arXiv:1405.6967 [hep-ph];
    N.~Chen, Z.~Liu and P.~Nath,
  arXiv:1406.0687 [hep-ph];
    J.~P.~Conlon and A.~J.~Powell,
  arXiv:1406.5518 [hep-ph];
  H.~Ishida and H.~Okada,
  arXiv:1406.5808 [hep-ph].
  

\bibitem{transition dipole}
  E.~Masso, S.~Mohanty and S.~Rao,
  Phys.\ Rev.\ D {\bf 80}, 036009 (2009)
  [arXiv:0906.1979 [hep-ph]].

\bibitem{anapole}
  Y.~Gao, C.~M.~Ho and R.~J.~Scherrer,
  Phys.\ Rev.\ D {\bf 89}, 045006 (2014)
  [arXiv:1311.5630 [hep-ph]].

\bibitem{oy}
  N.~Okada and T.~Yamada,
  JHEP {\bf 1310}, 017 (2013)
  [arXiv:1304.2962 [hep-ph]].

\bibitem{kolb turner}
 See, e.g., E.~W.~Kolb, and M.~S.~Turner, 
 {\it The Early Universe} (Addison-Wesley). 

\bibitem{Edsjo:1997bg} 
  J.~Edsjo and P.~Gondolo,
  Phys.\ Rev.\ D {\bf 56}, 1879 (1997)
  [hep-ph/9704361].

\bibitem{planck}
 P.~A.~R.~Ade \textit{et al.} (Planck Collaboration)
 [arXiv:1303.5076].

\bibitem{mt2}
 C.~G.~Lester and D.~J.~Summers,
 Phys.\ Lett.\ B {\bf 463}, 99 (1999)
 [hep-ph/9906349].

\bibitem{atlas}
 ATLAS collaboration,
 JHEP {\bf 05}, 071 (2014)
 [arXiv:1403.5294 [hep-ex]].

\bibitem{cms}
 CMS collaboration,
 arXiv:1405.7570 [hep-ex].

\bibitem{mg5}
 J.~Alwall et al., JHEP {\bf 0709}, 028 (2007) [arXiv:0706.2334[hep-ph]];
 J.~Alwall, M.~Herquet, F.~Maltoni, O.~Mattelaer and T.~Stelzer, 
 JHEP {\bf 1106}, 128 (2011) [arXiv:1106.0522[hep-ph]].

\bibitem{pythia}
 T.~Sjostrand, S.~Mrenna and P.~Z.~Skands, 
 JHEP {\bf 0605}, 026 (2006) [arXiv:hep-ph/0603175].

\bibitem{pgs4}
 J.~Conway et al., PGS (Pretty Good Simulation) (2009),
 http://physics.ucdavis.edu/~conway/research/software/pgs/pgs4-general.htm.
 
\bibitem{mlm}
 M.~L.~Mangano, M.~Moretti and R.~Pittau,
 Nucl.~Phys.~B {\bf 632}, 343 (2002) [arXiv:hep-ph/0108069];
 M.~L.~Mangano, M.~Moretti, F.~Piccinini and M.~Treccani,
 JHEP {\bf 0701}, 013 (2007) [arXiv:hep-ph/0611129].

\bibitem{antikt}
  M.~Cacciari, G.~P.~Salam and G.~Soyez,
  JHEP {\bf 0804}, 063 (2008)
  [arXiv:0802.1189 [hep-ph]].

\bibitem{Geng:2014zqa} 
  C.~-Q.~Geng, D.~Huang and L.~-H.~Tsai,
  arXiv:1406.6481 [hep-ph].


\end{thebibliography}
\end{document}